# Pan-genome Analysis of Angiosperm Plastomes using PGR-TK

Manoj P. Samanta

1. Systemix Institute, Sammamish, WA 98074.
2. Coding for Medicine LLC, Sammamish, WA 98074.
3. samanta@homolog.us

## Abstract

We present a novel approach for taxonomic analysis of chloroplast genomes in angiosperms using the Pan-genome Research Toolkit (PGR-TK). Comparative plots generated by PGR-TK across diverse angiosperm genera reveal a wide range of structural complexity, from straightforward to highly intricate patterns. Notably, the characteristic quadripartite plastome structure, comprising the large single copy (LSC), small single copy (SSC), and inverted repeat (IR) regions, is clearly identifiable in over 75% of the genera analyzed. Our findings also underscore several occurrences of species mis-annotations in public genomic databases, which are readily detected through visual anomalies in the PGR-TK plots. While more complex plot patterns remain difficult to interpret, they likely reflect underlying biological variation or technical inconsistencies in genome assembly. Overall, this approach effectively integrates classical botanical visualization with modern molecular taxonomy, providing a powerful tool for genome-based classification in plant systematics.

## Introduction

The advent of molecular biology and the development of DNA sequencing techniques in the late 20th century marked a paradigm shift in plant systematics [1-4]. Prior to this revolution, botanists relied primarily on morphological traits, such as the shape, size, and structure of flowers, leaves, stems, and fruits, to make informed judgments about vegetative and reproductive characteristics [5-6]. A landmark study by Chase *et al.* in 1993 highlighted the potential of molecular approaches by classifying approximately 500 seed plant species, representing all major taxonomic groups, using the rbcL gene from chloroplasts. This pioneering work laid the foundation for a new era in plant classification [1]. Subsequent studies built on this framework by incorporating additional genetic markers [7-9], further refining the understanding of plant relationships. Molecular phylogenetics has since enabled the construction of robust evolutionary trees, greatly enhancing the comprehension of plant diversity and evolutionary history.

While molecular classification offers a high degree of precision, current methodologies face several important limitations. Traditionally, taxonomic classification relied heavily on gene annotation. However, the rapid advancement of sequencing technologies [10], coupled with declining sequencing costs, led to an explosion in the availability of complete plastid genomes [11]. This surge presents new opportunities for whole-genome comparisons, including the often-neglected intergenic regions. These non-coding regions are particularly valuable for examining the biogeographical patterns of closely related species [12], where genetic differences may be subtle. Despite this promise, the comprehensive analysis of entire plastid genomes remains computationally demanding, a challenge further amplified by the increasing volume of data generated by next- and third-generation sequencing platforms. Moreover, the abstract and quantitative nature of molecular taxonomy can be alienating to traditional botanists. To bridge this divide, the development of intuitive visual tools that render molecular data in accessible, interpretable formats is essential. Such tools can facilitate the integration of genomic insights with classical taxonomy, ultimately enhancing the accuracy and accessibility of species identification.

The Pan-genome Research Toolkit (PGR-TK) offers an efficient solution to the challenges of analyzing complete plastid genomes without the need for gene annotations [13]. By leveraging a minimizer-based approach, PGR-TK significantly accelerates genome processing, enabling rapid and scalable comparisons of whole plastid genomes. In addition to its speed, the toolkit generates visual representations of similarity blocks, which help reveal evolutionary relationships among the species. It also incorporates principal bundle decomposition to analyze repeat structures within the genomes. The toolkit's effectiveness was demonstrated for four angiosperm genera [14], and this study extends its utility to assess genomic diversity across broader taxonomic scales, including genera, families, and orders.

# Results

Plastid genome sequences were downloaded from the NCBI database and organized into FASTA files, each containing sequences from a different angiosperm genus. PGR-TK was applied to 944 genera with three or more available sequences to generate visualizations of pan-genome conservation patterns. A parallel analysis was conducted at the family level, including all flowering plant families represented by at least three plastid genomes. To ensure consistency, all datasets were processed using the same set of PGR-TK parameters (see Methods). Circular genome sequences were standardized by aligning and rotating them to a common starting block identified by PGR-TK. Strand orientation was also normalized by reverse complementing sequences when necessary.

PGR-TK performs all-vs-all comparisons of the input sequences, identifying conserved segments shared across the genomes. It then produces visual plots in which each conserved region is represented as a colored rectangle, facilitating easy comparison of genome structure and conservation. In addition, PGR-TK generates a cladogram to illustrate evolutionary relationships among the sequences. For this study, each resulting profile was visually inspected to assess differences in sequence length and conservation patterns. The major trends observed across these plots are summarized below.

## Observations at the Genus level

**Pattern 1 - Clear Quadripartite Plastome Structures**

A typical chloroplast genome comprises a large single copy (LSC) region, a small single copy (SSC) region, and two inverted repeat (IR) segments that separate them. In over 75% of the genus-level plots generated by PGR-TK, these four distinct regions are clearly discernible (Fig. 1). The IR regions exhibit higher sequence conservation compared to the single copy regions, resulting in less fragmentation in these areas within the PGR-TK visualizations. To aid interpretation, PGR-TK highlights repeated segments using the same color and indicates their orientation with directional arrows (Fig. 1b). This feature is especially useful for examining the structure and symmetry of inverted repeats in chloroplast genomes. Finally, all sequences within the genus have nearly the same length in these plots.

From time to time, a subset of sequences within a genus exhibit an inverted SSC region relative to the majority. This likely reflects the presence of two chloroplast isoforms within cells, maintained in roughly equal proportions, that differ in the orientation of the SSC region relative to the LSC region. During genome assembly, only one of those isoforms is retained, and the choice appears to be somewhat random. The current study addressed this issue by inverting the SSC orientation in plastomes that differ from the majority, and subsequently reanalyzing the sequences using PGR-TK.

**Pattern 2 - Highly Fragmented Conserved Segments**

Excessive fragmentation is a notable pattern observed in a subset of the sequences. This indicates that the sequences exhibit high levels of divergence despite belonging to the same genus. Figure 2 displays a genus that exhibits a high degree of divergence, illustrating the extent of fragmentation within this group. Such fragmentation can complicate phylogenetic analyses and may suggest issues such as sequencing errors, misassembly, or biological factors like rapid evolution.

To systematically identify genera with fragmented sequences, two complementary approaches were employed: (i) PGR-TK plots were visually inspected to detect anomalies and breaks indicative of fragmentation, (ii) a metric calculating the number of PGR-TK segments per plastid was applied, providing an objective assessment of fragmentation levels. Both methods yielded consistent results. A list of 75 genera with highly fragmented segments is provided in Table 1.

**Pattern 3 - Misclassification of Plant Genus**

In a subset of plots, a single sequence displayed pronounced divergence from the rest. For example, in Fig. 3(a), the *Berchemia lineata* sequence clearly stands apart. To investigate this anomaly, a subsequence from the *Berchemia lineata* plastome was queried against the NCBI nucleotide database (NT) using BLAST, which revealed a strong match with *Phyllanthus* species. To further validate this finding, a combined PGR-TK plot of *Berchemia* and *Phyllanthus* was generated (Fig. 3(b)), which grouped *Berchemia lineata* with *Phyllanthus*, suggesting a likely misclassification.

Manual inspection of all PGR-TK plots identified 33 sets in which a single sequence (or two, in the case of *Ctenium*) showed marked divergence from the others. Each case was manually analyzed, with the findings summarized in Table 2. In two instances, the outlier species had already been removed from NCBI. Of the remaining cases, all but six were determined to involve misclassification of the chloroplast sequences. Additionally PGR-TK analysis across combined genera often accounted for the observed length differences in the outlier species.

**Pattern 4 - Short Sequences Primarily from Parasitic Plants**

While plastid genomes in most plants typically range from 140,000 to 160,000 nucleotides in length, a subset of species, primarily parasitic, exhibit substantially reduced plastid genomes. These reductions are generally associated with the loss of photosynthetic function. Broadly, angiosperm genera with short plastid sequences can be categorized into four distinct groups based on their length patterns and structural characteristics:

**i) Uniformly Short Sequences** *(e.g., Hydnora, Gastrodia, Yoania)*

In the genera *Hydnora* (order *Piperales*), *Gastrodia* and *Yoania* (order *Asparagales*), all available plastid sequences fall within the 40,000–60,000 nucleotide range. The PGR-TK plots for *Hydnora* and *Gastrodia* exhibit highly divergent structures (Fig. 5a, b), consistent with extensive plastome reduction and rearrangement. In contrast, *Yoania* retains a more canonical plastome organization, including intact inverted repeats (IRs) (Fig. 5c).

**ii) Short sequences with high length variability** *(e.g., Cuscuta, Orobanche)*

In *Cuscuta* and *Orobanche*, plastid genome lengths vary considerably among species. For instance, *Orobanche californica* diverges significantly in length and structure from the other *Orobanche* sequences. The PGR-TK plot for *Cuscuta* appears disordered, likely reflecting extensive gene loss and structural rearrangements across multiple species.

**iii) Predominantly standard-length sequences with one or two outliers** *(e.g., Burmannia, Solanum, Corybas, Neottia)*

In these genera, the majority of plastid genomes are of standard length and exhibit typical structural features. However, one or two sequences per genus deviate significantly, presenting with much shorter lengths. In each case, the standard-length sequences align well and conform to expected plastome structure, while the outlier(s) show poor alignment and irregular features.

**iv) Moderately reduced but structurally conserved plastomes** *(e.g., Zygophyllum)*

*Zygophyllum* represents an intermediate case. All plastid genomes analyzed are approximately 100,000 nucleotides in length, shorter than average but not extensively reduced. Despite the moderate reduction, PGR-TK plots for *Zygophyllum* display well-defined and conserved features, including distinct IRs, LSC and SSC regions, indicative of overall structural preservation.

**Pattern 5 - A Combination of Unconventional Patterns**

A small subset of the plots exhibited atypical patterns, characterized by mixed distributions of plastid lengths and notable structural variations. Three representative cases are presented in Fig. 5, with one examined in greater detail below:

(i) **_Cypripedium_** (order Asparagales): Within this orchid genus (Fig. 5a), four sequences, namely *Cypripedium × ventricosum*, *C. yunnanense*, *C. calceolus*, and *C. japonicum*, were highly similar in both length and structure. In contrast, the remaining sequences showed considerable divergence.

One sequence, *Cypripedium macranthos*, differed markedly from the others. Subsequent BLAST analysis revealed that this sequence belongs to *Hosta* (Fig. 6a) a genus in a different family within the same order, suggesting a potential misclassification or sample mix-up.

A reanalysis was conducted after excluding *C. macranthos* and applying partial or complete strand reversals to selected sequences. The resulting plot showed improved coherence and alignment across the dataset (Fig. 6b).

(ii) **_Rhododendron_** (order Ericales): Among 17 sequences analyzed, 11 displayed consistent structural patterns, while the remaining 6 diverged significantly in both length and structural features (Fig. 5b). Furthermore, these divergent sequences were also substantially larger. A comprehensive analysis is provided in the accompanying Ericales study.

(iii) **_Pelargonium_** (order Geraniales): The sequences exhibited pronounced differences in plastid length, indicating substantial structural variability within the genus (Fig. 5c).

## Observations at the Family levels

The comparative analysis using PGR-TK was extended to the family level by grouping sequences according to their respective families, each comprising one or more genera. As observed at the genus level, there was notable variation in plot complexity among families. In some orders, family-level plots revealed intricate and highly variable patterns, while in others, families exhibited consistently simple quadripartite structures. For example, all families within Liliales showed clear and well-resolved quadripartite plastome structures, whereas Caryophyllales members displayed a heterogeneous array of patterns. Several families within Caryophyllales exhibited highly complex configurations that diverged markedly from the canonical plastome organization. These findings suggest that in certain families, genomic divergence may be substantial enough to obscure the broader structural simplicity observed at higher taxonomic ranks.

To assess the resolution of PGR-TK at finer taxonomic scales, it was investigated whether individual genera within each family separated cleanly in the generated phylogeny. In general, this was indeed the case with genera typically forming distinct and coherent clusters.

Another noteworthy finding was the consistent visual distinctiveness of parasitic species characterized by highly reduced plastid genomes. These species often produced short, atypical sequences that stood out starkly in the plots, likely due to the loss of genes and structural regions commonly retained in autotrophic lineages.

To illustrate various family-level patterns in detail, four representative examples are provided in separate reports [15-18].

# Discussions

**Integrative Philosophy of PGR-TK**

Traditional methods of plastid sequence analysis face several inherent limitations that restrict their scalability and accessibility. Some of the most pressing challenges are computational demands to handle exponentially increasing sequence volume, the lack of ability to incorporate information from the entire genome and the lack of intuitive visual tools to interpret results. An integrated approach that combines the strengths of computational rigor with intuitive visual representations offers a promising solution. By fusing traditional botanical perspectives with modern algorithmic tools, researchers can gain both the analytical power and interpretability required for robust plastid genome analysis. This synergy can improve transparency, facilitate error detection, and enhance accessibility for users across disciplines.

PGR-TK embodies this integrative philosophy. It provides a visual, interactive platform that maintains the mathematical precision essential for high-quality plastid sequence analysis. One of its primary strengths lies in its ability to reveal conservation patterns across plastid genomes. Typically, plastid sequences are highly conserved within a genus, providing strong phylogenetic signals. However, exceptions do occur. For example, some plastids exhibit partial inversions affecting specific genomic regions. These anomalies are often difficult to detect using gene-based analysis but become visually evident through PGR-TK plots. Similarly, unusually long sequences, potentially the result of assembly or annotation errors, stand out clearly in the visual output.

Unlike traditional phylogenetic tools that often focus solely on coding regions, PGR-TK incorporates the entire plastid genome, including both coding and noncoding regions. This comprehensive approach enables the detection of deeper evolutionary relationships that might be overlooked in more limited analyses. These non-coding regions may also become valuable for examining the biogeographical patterns of closely related species, where genetic differences may be subtle.

**Errors in Plant Identification**

Although the majority of the plots displayed the characteristic quadripartite structure of the plastid genome, a small subset exhibited more complex and atypical patterns. It came as a surprise that some of these complex patterns were the results of plant misclassification. The application of PGR-TK proved instrumental in addressing this issue. When the PGR-TK analysis was conducted using representatives from both genera simultaneously, a notable correction occurred: the plastid genomes that had been erroneously annotated were clearly clustered with the genus to which they actually belonged. This realignment eliminated the previously observed visual distortions, thereby confirming the effectiveness of PGR-TK in identifying and correcting annotation discrepancies. Often this method also resolved issues regarding plastid length differences.

**Biological Significance of Other Complex Patterns**

In addition to technical errors, some plastid sequences exhibit genuinely complex patterns that deviate from standard phylogenetic expectations. These anomalies could represent rare or poorly understood evolutionary events, thereby offering valuable opportunities for novel biological insights. Recognizing and thoroughly investigating such irregularities is crucial for advancing our understanding of plant evolution.

It is important to note that the delineation of genera and families was originally based on morphological observations made over the last two centuries. Modern molecular phylogenetics largely preserved these classifications when they were monophyletic, shifting the focus toward resolving relationships among families and organizing them into orders and higher taxonomic ranks. As a result, the classification system now reflects a hybrid approach that integrates both morphological and molecular data. Consequently, genetic divergence within genera and families can be highly variable and is not uniformly distributed across groups.

The observations made through PGR-TK raise several important questions. Do nuclear genomes and their encoded genes in these highly divergent genera exhibit a similar degree of divergence? If so, could this imply underlying biochemical differences among the species within the same genus? Additionally, are some genera (e.g. *Xyris*) less rigorously defined because of their small plant and flower sizes? Should certain genera be split into multiple groups based on the observed chloroplast genome divergence?

**Genome Evolution**

PGR-TK is a powerful tool for exploring chloroplast genome evolution, capturing features such as plastid length variation, gene rearrangements, and changes in intergenic regions. While this study did not utilize the gene sequences directly, the tool supports their inclusion in the generated visualizations. Moreover, to support researchers who require access to sequence-level data, PGR-TK provides text files at every stage of the analysis. These outputs can be further processed or integrated into other workflows, promoting flexibility and reproducibility in genomic studies. Several steps in our analysis relied on these text files (see Methods). Finally, PGR-TK generates a dynamic, interactive interface that enhances both user engagement and data interpretability, functionality that cannot be fully captured through static images or written descriptions.

# Methods

## A. Raw Data

Plastid genomes from the RefSeq database were downloaded from NCBI (https://ftp.ncbi.nlm.nih.gov/refseq/release/plastid/). Using custom Biopython scripts, the files were split into FASTA format, with sequences organized by genus. Only files containing three or more sequences were retained for further analysis. Additionally, NCBI files were segregated into FASTA files for angiosperm families with three or more complete plastid genomes.

### B. PGR-TK Analysis

Once collected, PGR-TK was used to generate the alignments and compare different species in a genus or family. First, 'pgr-pbundle-decomp' was used with the following parameters to generate the conserved segments (principal bundles) in the sequences.

*pgr-pbundle-decomp -w 20 -k 32 -r 1 --min-branch-size 10 --min-span 0 --min-cov 0 --bundle-length-cutoff 10 extract2.fa out*

Next 'pgr-pbundle-bed2dist' was applied on the bundles to generate the phylogeny.

*pgr-pbundle-bed2dist out.bed ooo*

Finally, 'pgr-pbundle-bed2svg' was used to generate visual representations of the data.

*pgr-pbundle-bed2svg --ddg-file ooo.ddg --html out.bed plot*

### C. Rotation

A custom script was developed to align plastid sequences to a common origin. PGR-TK identified conserved blocks across sequences, and all plastids were rotated to align with this common block. The steps followed in the analysis were:

*pgr-pbundle-decomp -w 20 -k 32 -r 1 --min-branch-size 10 --min-span 0 --min-cov 0 --bundle-length-cutoff 10 $1 out*
*rotate.py $1 > extract2.fa*
*pgr-pbundle-decomp -w 20 -k 32 -r 1 --min-branch-size 10 --min-span 0 --min-cov 0 --bundle-length-cutoff 10 extract2.fa out*
*pgr-pbundle-bed2dist out.bed ooo*
*pgr-pbundle-bed2svg --ddg-file ooo.ddg --html out.bed plot*

### D. Inversion

A custom script was written to partially invert the LSC regions of specific plastid genomes. Principal bundles were used to identify inverted repeat regions, which were then replaced with their reverse complement.

### E. Density Analysis

Density analysis was conducted on the bed file generated by 'pgr-pbundle-decomp'. This analysis counted the number of principal bundles for each plastid. Since plastid lengths within a genus typically do not vary significantly, no normalization was required.

### F. BLAST Analysis of Incorrectly Annotated Sequences

A subset of plastid genomes was identified as being incorrectly annotated. BLAST was first used to search a subsequence from the outlier plastid against NCBI. The top hits, excluding the original plastid, were examined to suggest potential alternate genera. If a match was found, a combined PGR-TK analysis of both genera was performed to visually assess the accuracy of the BLAST annotation.

# Figures

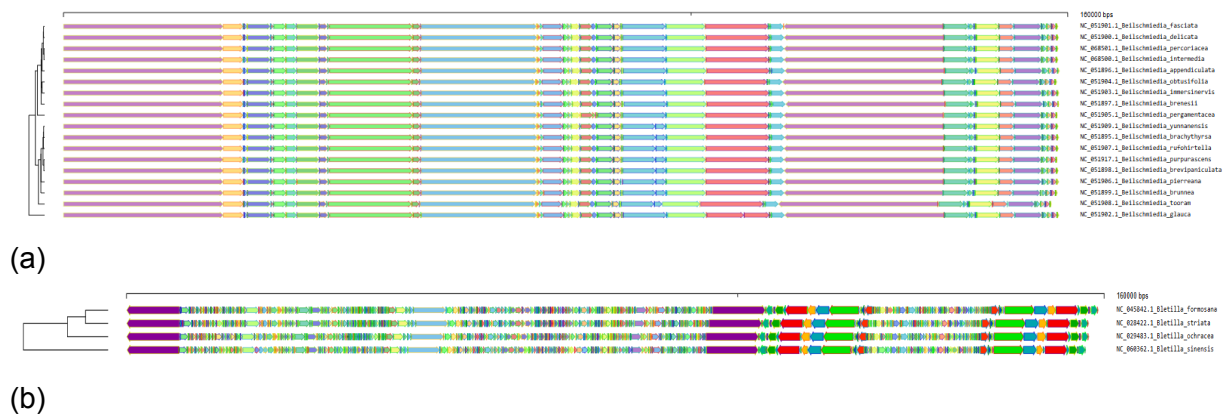

**Figure 1. PGR-TK plots for (a) *Beilschmiedia* and (b) *Bletilla*.** These plots display the common pattern followed by over 75% of the 944 analyzed genera. Four regions of the chloroplast, namely long single copy (LSC), short single copy (SSC) and the pair of inverted repeat (IR), are clearly visible. The inverted repeat segments are highlighted in Fig. 1(b) using PGR-TK's interactive visualization tool.

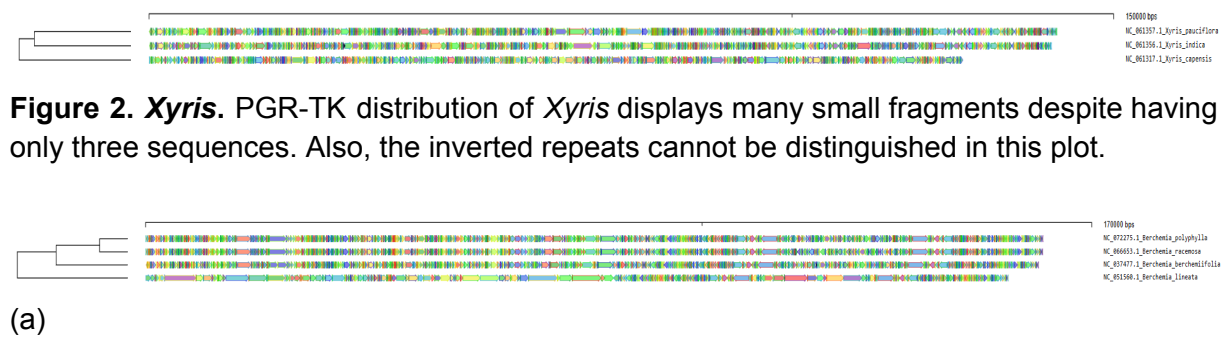

**Figure 2. *Xyris*.** PGR-TK distribution of *Xyris* displays many small fragments despite having only three sequences. Also, the inverted repeats cannot be distinguished in this plot.

(a)

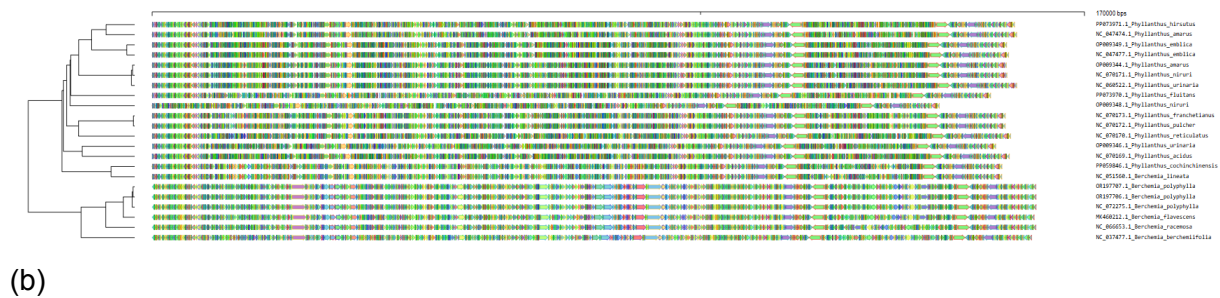

(b)

**Figure 3. PGR-TK plots for (a)** *Berchemia* **and (b)** *Berchemia* **and** *Phyllantha* **combined.** In (a), *Berchemia lineata* shows a distribution completely different from the remaining sequences. When the PGR-TK plot of *Berchemia* and *Phyllantha* are drawn together (b), *Berchemia lineata* aligns with Phyllantha.

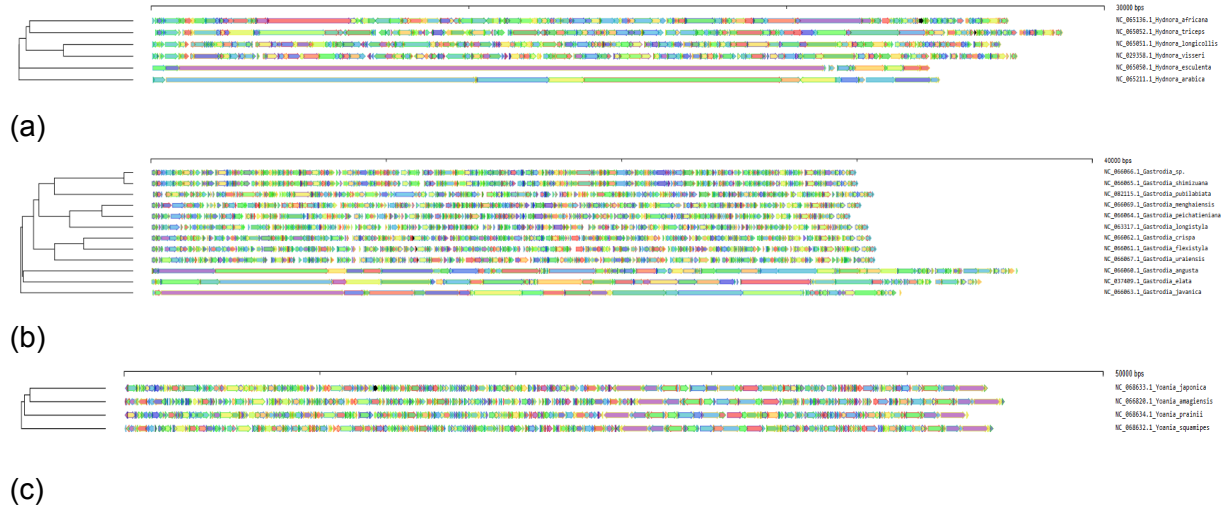

(a)

(b)

(c)

**Figure 4. PGR-TK plots for (a)** *Hydnora*, **(b)** *Gastrodia* **and (c)** *Yoania*. The parasitic plants *Hydnora* and *Gastrodia* show significant divergence in PGR-TK distributions, whereas *Yoania*, despite having small genomes, shows clear signatures of inverted repeats.

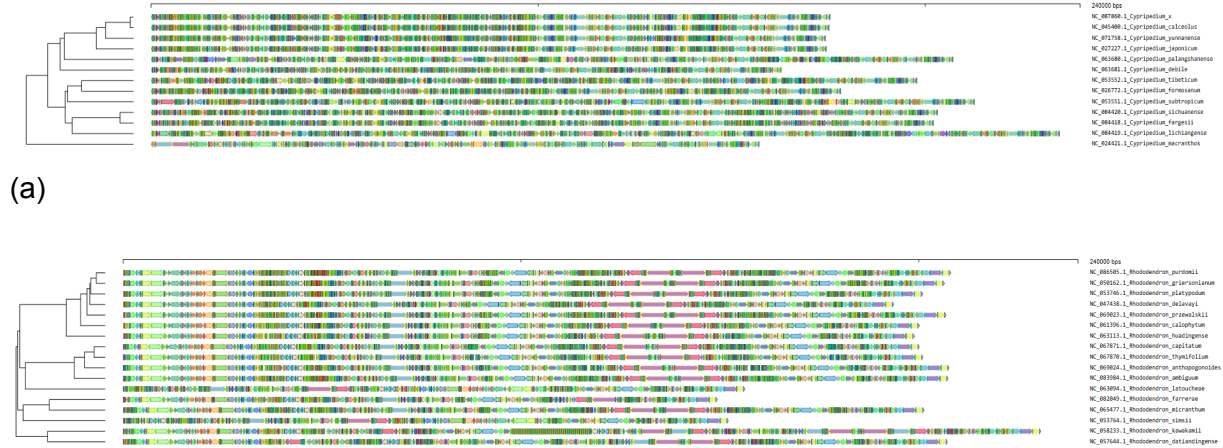

(a)

(b)

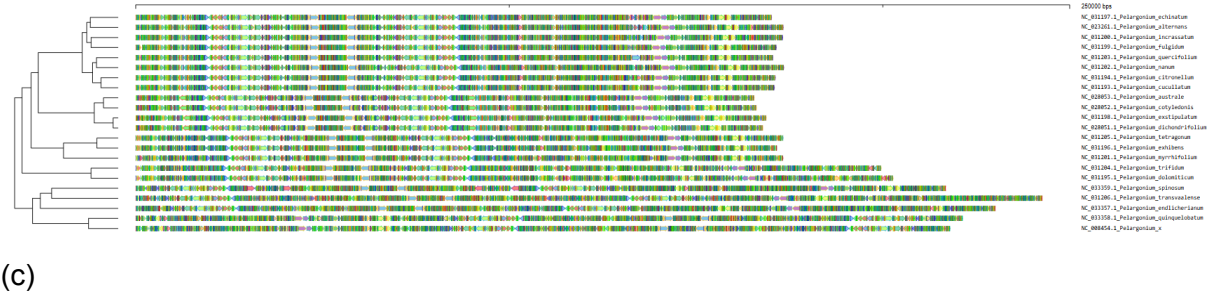

(c)

**Figure 5. PGR-TK plots for (a) *Cypripedium*, (b) *Rhododendron* and (c) *Pelargonium*.** Plastomes in these plots show varied lengths and dissimilarity between the sequences.

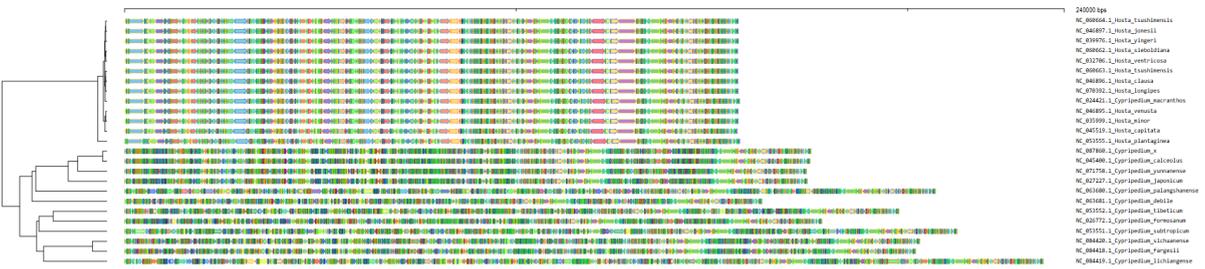

(a)

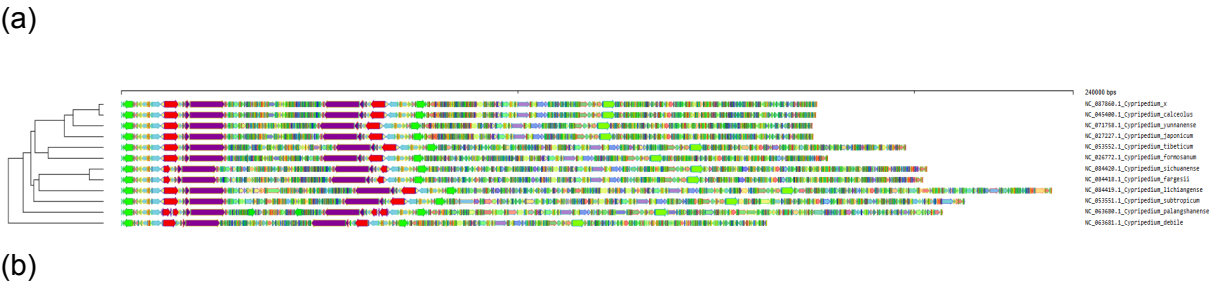

(b)

**Figure 6. PGR-TK plots for (a) *Cypripedium* and *Hosta* together, (b) *Cypripedium* (modified).** In Fig. (a) *Cypripedium macranthos* aligns with *Hosta*, and therefore is mis-annotated. Fig. (b) is drawn after excluding *Cypripedium macranthos* and reorienting other genomes.

# Tables

| Order | Genus | Comment |
|---|---|---|
| Alismatales | *Hydrocharis* | Highly fragmented, no clear match between species |
| Alismatales | *Zantedeschia* | Highly fragmented, no clear match between species |
| Dioscoreales | *Burmannia* | Varied length and highly fragmented |
| Asparagales | *Corallorhiza* | Highly fragmented, but shows strong match between species. Two sequences are shorter than the rest. |
| Asparagales | *Corybas* | *Corybas cryptanthus* much shorter than the rest |
| Asparagales | *Cypripedium* | Highly fragmented, multiple lengths |
| Asparagales | *Neottia* | Varied length and highly fragmented |
| Asparagales | *Vanilla* | Highly fragmented, but shows strong match between species |
| Poales | *Carex* | Highly fragmented, no clear match between species |
| Poales | *Ctenium* | Two sequences substantially different from the rest |
| Poales | *Juncus* | Highly fragmented, no clear match between species |
| Poales | *Xyris* | Highly fragmented, no clear match between species |
| Ranunculales | *Corydalis* | Many sequences, complex patterns |
| Ranunculales | *Helleborus* | Varied lengths and sequence patterns |
| Ranunculales | *Hypecoum* | Highly fragmented except in the inverted repeats |
| Ranunculales | *Semiaquilegia* | High fragmentation despite having three sequences |
| Saxifragales | *Crassula* | Highly fragmented, but shows strong match between species. Two sequences are substantially different. |
| Oxalidales | *Oxalis* | Highly fragmented except in the inverted repeats |
| Malpighiales | *Euphorbia* | Many sequences, complex patterns |
| Malpighiales | *Hypericum* | Highly fragmented |
| Malpighiales | *Linum* | Highly fragmented |
| Fabales | *Bauhinia* | Highly fragmented |
| Fabales | *Hedysarum* | Highly fragmented |
| Fabales | *Lathyrus* | Highly fragmented |
| Fabales | *Medicago* | Many sequences, complex patterns |
| Fabales | *Mucuna* | Highly fragmented |
| Fabales | *Polygonum* | Highly fragmented |
| Fabales | *Trifolium* | Highly fragmented |

| Fabales | *Vicia* | Highly fragmented |
|---|---|---|
| Rosales | *Berchemia* | Highly fragmented |
| Rosales | *Boehmeria* | Highly fragmented |
| Rosales | *Laportea* | Highly fragmented except in the inverted repeats |
| Rosales | *Pilea* | Highly fragmented except in the inverted repeats |
| Cucurbitales | *Begonia* | Many sequences, complex patterns |
| Geraniales | *Erodium* | Highly fragmented |
| Geraniales | *Geranium* | Highly fragmented |
| Geraniales | *Monsonia* | Highly fragmented |
| Geraniales | *Pelargonium* | Highly fragmented, difference in lengths |
| Myrtales | *Eugenia* | Highly fragmented |
| Brassicales | *Chorispora* | Highly fragmented except in inverted repeats |
| Santalales | *Scurrula* | Highly fragmented |
| Santalales | *Viscum* | Highly fragmented |
| Caryophyllales | *Arenaria* | Highly fragmented |
| Caryophyllales | *Drosera* | Highly fragmented |
| Caryophyllales | *Salsola* | Highly fragmented |
| Caryophyllales | *Silene* | Complex pattern with many sequences |
| Caryophyllales | *Suaeda* | Highly fragmented |
| Ericales | *Androsace* | Highly fragmented except in inverted repeats |
| Ericales | *Gordonia* | Highly fragmented |
| Ericales | *Impatiens* | Highly fragmented except in inverted repeats |
| Ericales | *Rhododendron* | Highly fragmented, many lengths |
| Ericales | *Vaccinium* | Fragmented and variations in lengths |
| Boraginales | *Heliotropium* | Highly fragmented |
| Gentianales | *Gynochthodes* | Highly fragmented |
| Gentianales | *Swertia* | Highly fragmented except in inverted repeats |
| Lamiales | *Acanthus* | Highly fragmented |
| Lamiales | *Barleria* | Highly fragmented |
| Lamiales | *Echinacanthus* | Highly fragmented |
| Lamiales | *Genlisea* | Highly fragmented |
| Lamiales | *Incarvillea* | Highly fragmented |
| Lamiales | *Orobanche* | Highly fragmented and many lengths |
| Lamiales | *Pedicularis* | Complex patterns with many sequences |
| Lamiales | *Utricularia* | Combination of many patterns, highly fragmented |
| Solanales | *Cuscuta* | Many lengths, highly fragmented |
| Solanales | *Solanum* | 3-4 sequences are short |
| Asterales | *Adenophora* | Highly fragmented |
| Asterales | *Campanula* | Highly fragmented |
| Asterales | *Codonopsis* | Highly fragmented |
| Asterales | *Cyphia* | Highly fragmented |
| Asterales | *Lobelia* | Complex pattern with many sequences |
| Apiales | *Pimpinella* | Highly fragmented |
| Apiales | *Pleurospermum* | Highly fragmented |
| Apiales | *Sinocarum* | Highly fragmented |
| Apiales | *Tongoloa* | Highly fragmented |
| Apiales | *Trachydium* | Highly fragmented |

**Table 1. PGR-TK plots displaying complex patterns.** These 75 angiosperm genera exhibit highly fragmented or complex PGR-TK patterns, often accompanied by substantial variation in plastid genome lengths within each genus.

| Original Annotation | Revised Annotation | Comment |
|---|---|---|
| *Adenophora triphylla* | X | NCBI currently has two sequences for Adenophora triphylla, and they are highly divergent. Also, both sequences are different from the remaining Adenophora sequences. Most likely, this sequence has sequencing or assembly error. |
| *Androsace septentrionalis* | X | This sequence is likely *Androsace*. |
| *Barleria siamensis* | *Strobilanthes crispa* | Revised annotation is confirmed by BLAST and PGR-TK. Also, revision explains the length difference. |
| *Bauhinia binata* | *Lysiphyllum hookeri* | Revised annotation is confirmed by BLAST and PGR-TK. Also, revision is consistent with the minor length difference. |

| Original Annotation | Revised Annotation | Comment |
|---|---|---|
| *Adenophora triphylla* | X | NCBI currently has two sequences for Adenophora triphylla, and they are highly divergent. Also, both sequences are different from the remaining Adenophora sequences. Most likely, this sequence has sequencing or assembly error. |
| *Androsace septentrionalis* | X | This sequence is likely *Androsace*. |
| *Barleria siamensis* | *Strobilanthes crispa* | Revised annotation is confirmed by BLAST and PGR-TK. Also, revision explains the length difference. |
| *Begonia pedatifida* | X | Sequence is removed from NCBI. |
| *Berchemia lineata* | *Phyllanthus cochinchinensis* | Revised annotation is confirmed by BLAST and PGR-TK. Also, revision explains the length difference. |
| *Cinnamomum longipetiolatum* | *Beilschmiedia pergamentacea* | Revised annotation is confirmed by BLAST and PGR-TK. Also, revision explains the length difference. |
| *Codonopsis convolvulacea* | *Pseudocodon vinciflorus* | Revised annotation is confirmed by BLAST and PGR-TK. Also, the revision explains the length difference. |
| *Corydalis bungeana* | *Viola japonica* | Not checked using PGR-TK. |
| *Ctenium aromaticum* | X | Nearest BLAST matches are all from *Ctenium.* |
| *Ctenium cirrhosum* | X | Sequence removed from NCBI. |
| *Cyclamen persicum* | *Pseudostellaria japonica* | BLAST and PGR-TK confirm the new annotation. |
| *Cypripedium macranthos* | *Hosta longipes* | Revised annotation is confirmed by BLAST and PGR-TK. Also, the revision explains the length difference. |
| *Dendrobium trigonopus* | *Lonchocarpus utilis* | BLAST and PGR-TK confirm the new annotation. |
| *Elymus antiquus* | *Agrostis stolonifera* | BLAST and PGR-TK confirm the new annotation. |
| *Eugenia brasiliensis* | *Aspidopterys glabriuscula* | BLAST and PGR-TK confirm the new annotation. |
| *Genlisea violacea* | X | BLAST top hits include *Utricularia volubilis,* but PGR-TK does not confirm. |

| Original Annotation | Revised Annotation | Comment |
|---|---|---|
| *Adenophora triphylla* | X | NCBI currently has two sequences for Adenophora triphylla, and they are highly divergent. Also, both sequences are different from the remaining Adenophora sequences. Most likely, this sequence has sequencing or assembly error. |
| *Androsace septentrionalis* | X | This sequence is likely *Androsace*. |
| *Barleria siamensis* | *Strobilanthes crispa* | Revised annotation is confirmed by BLAST and PGR-TK. Also, revision explains the length difference. |
| *Gynochthodes nanlingensis* | *Epigynum auritum* | Not checked using PGR-TK. |
| *Heritiera elata* | *Diospyros maclurei* | BLAST and PGR-TK confirm the new annotation. |
| *Ipomoea setifera* | *Durio graveolens* | Weak BLAST match. PGR-TK clearly clusters with Durio, but shows sequence differences. |
| *Lagerstroemia villosa* | *Terminalia guyanensis* | 2 Lagerstroemia villosa - different. BLAST and PGR-TK confirm the new annotation. |
| *Lobelia thermalis* | *Monopsis flava* | Not checked using PGR-TK. |
| *Microtropis biflora* | *Adina rubella* | Not checked using PGR-TK. |
| *Orobanche californica* | *Brandisia glabrescens* | Weak blast match |
| *Ostericum grosseserratum* | *Angelica tianmuensis* | Revised annotation is confirmed by BLAST and PGR-TK. Also, the revision explains the length difference. |
| *Pedicularis cheilanthifolia* | *Dracocephalum heterophyllum* | Weak blast match |
| *Pilea penninervis* | *Elatostema stewardii* | BLAST and PGR-TK confirm the new annotation. |
| *Potentilla sischanensis* | X | Nearest BLAST matches are all from *Potentilla*. |
| *Prunus kanzakura* | X | Nearest BLAST matches are all from *Prunus*. |
| *Silene noctiflora* | X | Nearest BLAST matches are all from *Silene*. |
| *Sinocarum* | *Sinolimprichtia sp.* | Not confirmed by PGR-TK. *Sinocarum* |

| Original Annotation | Revised Annotation | Comment |
|---|---|---|
| *Adenophora triphylla* | X | NCBI currently has two sequences for Adenophora triphylla, and they are highly divergent. Also, both sequences are different from the remaining Adenophora sequences. Most likely, this sequence has sequencing or assembly error. |
| *Androsace septentrionalis* | X | This sequence is likely *Androsace*. |
| *Barleria siamensis* | *Strobilanthes crispa* | Revised annotation is confirmed by BLAST and PGR-TK. Also, revision explains the length difference. |
| *schizopetalum* | | *schizopetalum* appears intermediate between two genera. |
| *Tongoloa zhongdianensis* | *Angelica ternata* | Both BLAST and PGR-TK show that *Tongoloa zhongdianensis* should be assigned to the *Angelica* genus. |
| *Viscum articulatum* | *Korthalsella sp.* | PGR-TK analysis suggests that *Viscum articulatum* should be reassigned to the *Korthalsella* genus. This is likely to be a taxonomic reassignment, and not an error in sample annotation. |
| *Zephyranthes bifida* | *Dioscorea futschauensis* | Both BLAST and PGR-TK show that *Zephyranthes bifida* should be assigned to the *Dioscorea* genus. Given the evolutionary distance between the current and revised annotation, the current assignment is likely an error. |

**Table 2. Angiosperm genera with one sequence diverging from the rest.** Among the 33 angiosperm genera listed in the table above, one species in each genus (two in the case of *Ctenium*) diverges from the others. Further analysis using BLAST and PGR-TK revealed that, in most cases, the divergent species had been incorrectly annotated.